\documentclass[sigconf,screen]{acmart}

\AtBeginDocument{%
  \providecommand\BibTeX{{%
    \normalfont B\kern-0.5em{\scshape i\kern-0.25em b}\kern-0.8em\TeX}}}

\copyrightyear{2024}
\acmYear{2024}
\setcopyright{acmlicensed}\acmConference[ICSE '24]{2024 IEEE/ACM 46th International Conference on Software Engineering}{April 14--20, 2024}{Lisbon, Portugal}
\acmBooktitle{2024 IEEE/ACM 46th International Conference on Software Engineering (ICSE '24), April 14--20, 2024, Lisbon, Portugal}
\acmDOI{10.1145/3597503.3639100}
\acmISBN{979-8-4007-0217-4/24/04}

\usepackage{amsmath,amssymb,amsfonts}
\usepackage{graphicx}
\usepackage{textcomp}
\usepackage{color}
\usepackage{adjustbox}
\usepackage{makecell}
\usepackage{multirow}
\usepackage{threeparttable}
\usepackage{enumitem}
\usepackage{booktabs}
\usepackage{bbding}
\usepackage{colortbl}
\usepackage{xcolor}
\usepackage{setspace}
\usepackage{pifont}
\usepackage{caption}
\usepackage{subcaption}
\usepackage{url}
\usepackage{fontawesome}
\usepackage{listings}
\usepackage[ruled,linesnumbered,vlined]{algorithm2e}
\usepackage[noend]{algpseudocode}
\usepackage[most]{tcolorbox}

\newcommand{\parabf}[1]{\noindent\textbf{#1}}
\newcommand{\codeIn}[1]{\texttt{\small #1}}

\newcommand{\mybox}[1]{\begin{tcolorbox}[enhanced, frame hidden, boxsep=0pt]\emph{#1}\end{tcolorbox}}

\definecolor{codeblue}{RGB}{0, 119, 182}
\lstdefinestyle{myStyle}{ 
    commentstyle=\color{codeblue},
    basicstyle=\ttfamily\scriptsize,
    breakatwhitespace=false,
    breaklines=true,
    frame=lines,
    keepspaces=false,                 
    numbers=none,       
    numbersep=5pt,                  
    showspaces=false,                
    showstringspaces=false,
    showtabs=false,                  
    tabsize=2
}
\newcommand{\binaryai}{BinaryAI}
\newcommand{\osspolice}{OSSPolice}
\newcommand{\bsfinder}{B2SFinder}
\newcommand{\binpro}{BinPro}
\newcommand{\bdba}{bsca-B}
\newcommand{\scantist}{bsca-A}
\newcommand{\linklocality}{link-time locality}
\newcommand{\linklocalities}{link-time localities}
\newcommand{\fcg}{function call graph}
\newcommand{\ApproachFeature}{Feature Extraction}
\newcommand{\approachfeature}{feature extraction}
\newcommand{\ApproachEmbedding}{Embedding-based Function Retrieval}
\newcommand{\approachembedding}{\textit{embedding-based function retrieval}}
\newcommand{\ApproachLocality}{Locality-driven Matching}
\newcommand{\approachlocality}{\textit{locality-driven matching}}
\newcommand{\ApproachSCA}{Third-party Library Detection}
\newcommand{\approachsca}{\textit{third-party library detection}}
\newcommand{\tplite}{TPLite}
\newcommand{\codecmr}{CodeCMR}

\newcommand{\tplnum}{12,013}
\newcommand{\srcfuncnum}{56,342,179}
\newcommand{\scaproduct}{85}
\newcommand{\scabin}{150}
\newcommand{\scalabel}{1,045}
\newcommand{\scare}{15}
\newcommand{\scarefunc}{23,529}
\newcommand{\testfunc}{32,296}
\newcommand{\totalpair}{10M}

\newcommand{\basemodel}{{{\textit{Pythia}}}}
\newcommand{\binaryaimrr}{0.34}
\newcommand{\codecmrmrr}{0.17}

\begin{document}
\title{\binaryai{}: Binary Software Composition Analysis via Intelligent Binary Source Code Matching}

\author{Ling Jiang}
\affiliation{%
  \institution{Research Institute of Trustworthy Autonomous Systems, Southern University of Science and Technology}
  \city{Shenzhen}
  \country{China}
}
\email{11711906@mail.sustech.edu.cn}

\author{Junwen An}
\affiliation{%
  \institution{Southern University of Science and Technology}
  \city{Shenzhen}
  \country{China}
}
\email{12012109@mail.sustech.edu.cn}

\author{Huihui Huang}
\affiliation{%
  \institution{Southern University of Science and Technology}
  \city{Shenzhen}
  \country{China}
}
\email{12010336@mail.sustech.edu.cn}

\author{Qiyi Tang, Sen Nie, Shi Wu}
\affiliation{%
  \institution{Tencent Security Keen Lab}
  \city{Shanghai}
  \country{China}
}
\email{{dodgetang, snie, shiwu}@tencent.com}




\author{Yuqun Zhang}
\authornote{Yuqun Zhang is the corresponding author. He is also affiliated with the Department of Computer Science and Engineering, Southern University of Science and Technology, Shenzhen, China and Guangdong Provincial Key Laboratory of Brain-inspired Intelligent Computation, China.}
\affiliation{%
  \institution{Research Institute of Trustworthy Autonomous Systems, Southern University of Science and Technology}
  \city{Shenzhen}
  \country{China}
}
\email{zhangyq@sustech.edu.cn}

\ccsdesc[300]{Security and privacy~Software security engineering}
\ccsdesc[300]{Software and its engineering~Software libraries and repositories}

\keywords{Software Composition Analysis, Static Binary Analysis}

\begin{abstract}

While third-party libraries (TPLs) are extensively reused to enhance productivity during software development, they can also introduce potential security risks such as vulnerability propagation. Software composition analysis (SCA), proposed to identify reused TPLs for reducing such risks, has become an essential procedure within modern DevSecOps.
As one of the mainstream SCA techniques, binary-to-source SCA identifies the third-party source projects contained in binary files via binary source code matching, which is a major challenge in reverse engineering since binary and source code exhibit substantial disparities after compilation. The existing binary-to-source SCA techniques leverage basic syntactic features that suffer from redundancy and lack robustness in the large-scale TPL dataset, leading to inevitable false positives and compromised recall. To mitigate these limitations, we introduce \binaryai{}, a novel binary-to-source SCA technique with two-phase binary source code matching to capture both syntactic and semantic code features. First, \binaryai{} trains a transformer-based model to produce function-level embeddings and obtain similar source functions for each binary function accordingly. Then by applying the \linklocality{} to facilitate function matching, \binaryai{} detects the reused TPLs based on the ratio of matched source functions. Our experimental results demonstrate the superior performance of \binaryai{} in terms of binary source code matching and the downstream SCA task. Specifically, our embedding model outperforms the state-of-the-art model \codecmr{}, i.e., achieving 22.54\% \textit{recall@1} and \binaryaimrr{} \textit{MRR} compared with 10.75\% and \codecmrmrr{} respectively. Additionally, \binaryai{} outperforms all existing binary-to-source SCA tools in TPL detection, increasing the precision from 73.36\% to 85.84\% and recall from 59.81\% to 64.98\% compared with the well-recognized commercial SCA product.

\noindent \includegraphics[height=0.25cm]{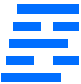} \href{https://www.binaryai.net}{ \textbf{https://www.binaryai.net}} 


\end{abstract}

\maketitle

\section{Introduction}

Software composition analysis (SCA)~\cite{tplite,yuan2019b2sfinder,blackduck} refers to identifying the open-source components (reused third-party libraries, i.e., TPLs) contained in the software artifacts for cost-effective development. Based on the SCA result, developers can easily track the security threats introduced to the software artifact by TPLs, such as vulnerability propagation and license violation~\cite{whatssca, sca_mend, sca_geekflare, Liu2022DemystifyingTV}. Considering diverse forms of target software project and identified TPLs, the existing SCA techniques are divided into multiple categories (e.g., binary-to-binary SCA~\cite{tang2020libdx, yang2022modx, tang2022libdb} and binary-to-source SCA~\cite{duan2017identifying, yuan2019b2sfinder, miyani2017binpro}). In particular, as one of the mainstream SCA techniques, binary-to-source SCA techniques identify the source code projects as reused TPLs contained in the target binary file by measuring the similarity between the binary file and a large-scale collected TPL dataset based on their extracted code features (i.e., binary source code matching). During the modern DevSecOps CI/CD pipeline, binary-to-source SCA is normally integrated into the build or deploy phase to automatically scan the components along with security risks within binary files. Typically, binary-to-source SCA tends to be scalable and practical in real-world software development scenarios~\cite{blackduck,duan2017identifying,supply-chain} by conveniently incorporating new open source repositories into the TPL dataset.

The existing binary-to-source SCA techniques utilize basic syntactic features (e.g., string literals) that remain consistent after compilation to perform binary source code matching. 
For instance, \bsfinder{}~\cite{yuan2019b2sfinder}, the state-of-the-art academic SCA technique, leverages a weighted matching algorithm to combine three matching techniques for the corresponding basic syntactic features. Although basic features can be used to build the correspondence between binary and source code, the existing binary-to-source SCA techniques relying on basic features still have two limitations. First, these basic features typically exhibit a significant degree of redundancy in the large-scale TPL dataset. Such an issue causes feature duplication across collected TPLs that can further limit the precision of SCA, i.e., incurring inevitable false positives during feature matching. Moreover, it is commonly observed that few or even no common basic syntactic features exist between reused TPLs and target binary files, especially the binary files which are stripped of certain string literals such as function names~\cite{yuan2019b2sfinder}, compromising the recall of SCA techniques~\cite{duan2017identifying} that rely on such features. In fact, binary code differs significantly from the source code, and few features remain consistent after compilation. Such disparity can even increase false negatives when applying traditional techniques to match basic features. Therefore, it is essential to employ finer-grained features, such as function-level features which typically contain more high-level syntactic and semantic information compared with basic syntactic features, to advance the accuracy of binary source code matching and further strengthen the downstream binary-to-source SCA task. 

In this paper, we propose \binaryai{}, a binary-to-source SCA technique with intelligent function-level binary source code matching. Given the disparities introduced by compilation, we adopt a transformer-based model to capture the token-based syntactic features and generate function embeddings for computing the similarity between binary and source functions. Specifically, \binaryai{} uses the large language model from the suite of \basemodel{}~\cite{biderman2023pythia} as the starting point, followed by pre-training the model in a supervised manner using contrastive learning~\cite{radford2021learning}. 
Based on the trained model, the embeddings for all source functions from a large-scale TPL dataset are generated offline and stored in a corpus (i.e., vector database). For the online SCA detection of the target binary file, \binaryai{} performs decompilation to extract binary functions, which are further encoded into embeddings as queries to retrieve \textit{top-k} similar source functions from the corpus. Furthermore, \binaryai{} adopts \approachlocality{} at the second phase of binary source function matching. Specifically, we leverage \linklocality{} and \fcg{} as additional structured information to capture the semantic features and identify the exactly matched source function from the \textit{top-k} similar functions. 
Eventually, \binaryai{} leverages the matched source functions to calculate the ratio of reused functions as the similarity score between collected TPLs and the target binary file, further identifying the components along with potential security risks whose similarity exceeds a pre-defined threshold as in many previous works~\cite{tang2022libdb, yuan2019b2sfinder, duan2017identifying, tang2020libdx}.


In this paper, we evaluate the effectiveness of \binaryai{} in terms of binary source code matching and TPL detection (i.e., SCA). Specifically, we first construct three datasets: 1) training set for the model containing around \totalpair{} function pairs as positive samples, 2) large-scale TPL dataset to construct the SCA database and corpus for retrieving similar source functions, and 3) SCA test set with manually labeled components and binary-to-source function mappings. The evaluation results reveal that the binary source code matching model in \binaryai{} outperforms the state-of-the-art model \codecmr{} by increasing \textit{recall@1} from 11.92\% to 22.73\% for the SCA test set. Moreover, the \approachlocality{} can effectively identify the correct source function from \textit{top-k} retrieved results, further increasing \textit{recall@1} from 22.73\% to 66.90\% that is close to the upper bound 70.45\% (i.e., \textit{recall@100}) restricted by the model capability. Based on the matched source functions, we evaluate the accuracy of \binaryai{} regarding TPL detection. The evaluation results demonstrate that \binaryai{} dominates the performance among all the existing binary-to-source SCA tools, e.g.,  outperforming the start-of-the-art academic SCA tool \bsfinder{} by increasing the precision from 31.78\% to 85.84\% and the recall from 54.93\% to 64.98\%. It even outperforms the well-recognized commercial SCA product by increasing the precision from 73.36\% to 85.84\% and the recall from 59.81\% to 64.98\%.

To summarize, our paper makes the following contributions:

\begin{itemize}[leftmargin=*]
    \item To our best knowledge, we are the first to adapt function-level binary source code matching to binary-to-source SCA and train a transformer-based model to retrieve similar source functions.

    \item We propose a two-phase binary source function matching in \binaryai{} by leveraging \linklocality{} to enhance the accuracy of function matching with the \textit{top-k} retrieved results.

    \item We evaluate \binaryai{}, where the results suggest that the model of \binaryai{} significantly outperforms \codecmr{} in binary source code matching. In addition, \binaryai{} dominates the performance among the existing binary-to-source SCA tools.
\end{itemize}
\section{Background and Motivation}

\subsection{Software Composition Analysis}
Software composition analysis (SCA) typically refers to identifying third-party libraries (TPLs) in the target software project to track security threats and license violations introduced by these open-source components. Given the potential risks to the software supply chain associated with accessing the source code (e.g., privacy policy), binary SCA has emerged as the predominant technique, which can be easily integrated into the build or deploy phase during DevSecOps to automatically scan the components in binary files~\cite{duan2017identifying, yuan2019b2sfinder, blackduck}. Existing binary SCA techniques~\cite{duan2017identifying, yuan2019b2sfinder, tang2020libdx, li2017libd, yang2022modx} extract software features from a large-scale TPL dataset to construct the SCA database and then utilize code clone detection to identify similar features between TPLs and the binary file. Subsequently, they recognize the TPLs as the reused components if the number of similar features exceeds a pre-defined threshold. Based on different forms of TPLs in the database, binary SCA can be divided into two categories: binary-to-source SCA~\cite{duan2017identifying, yuan2019b2sfinder, blackduck} and binary-to-binary SCA~\cite{tang2020libdx, tang2022libdb, yang2022modx}.

\subsubsection{Binary-to-Source SCA}
The TPL dataset in binary-to-source SCA consists of large-scale crawled open-source C/C++ projects, the majority of which are GitHub repositories and source packages from the GNU/Linux community. By matching source code features extracted from the C/C++ repositories, binary-to-source SCA identifies the reused source-code-level TPLs in the target binary file. Specifically, the fundamental step in binary-to-source SCA is binary source code matching, which maps binary code to the corresponding source code. \bsfinder{}~\cite{yuan2019b2sfinder}, the start-of-the-art tool, selects basic syntactic features (e.g., string literals) that still remain consistent after compilation to match the source code and  open-source components. In addition to binary SCA, binary source code matching is crucial in other scenarios of software security, such as reverse engineering~\cite{miyani2017binpro} and malware analysis~\cite{hemel2011finding}.  To our best knowledge, the effectiveness of existing binary source code matching is generally compromised due to substantial disparities between binary and source code~\cite{yu2020codecmr}.

\subsubsection{Binary-to-Binary SCA}
In the binary-to-binary SCA task, the TPLs in the SCA database are stored in the binary format built from source packages. By leveraging publicly available package managers (e.g., Nix~\cite{nix}), source packages can be compiled automatically across different versions, architectures, and optimization levels. Many existing binary-to-binary SCA techniques~\cite{tang2022libdb,yang2022modx,Li2023LibAMAA} integrate advanced embedding-based approaches to detect code similarity between binaries and further identify the reused libraries based on the SCA database. Specifically, they leverage deep neural network models 
to embed binary functions into the representation of vectors and perform binary code clone detection by measuring the similarity between function embeddings~\cite{xu2017neural,wang2022jtrans,ding2019asm2vec, marcelli2022howML}. Apart from basic syntactic features, these techniques typically capture semantic features such as the control flow graph (CFG) for each binary function to strengthen their accuracy 
of code clone detection and the downstream SCA task.

\begin{figure*}[t]
    \centering
    \includegraphics[width=0.93\textwidth]{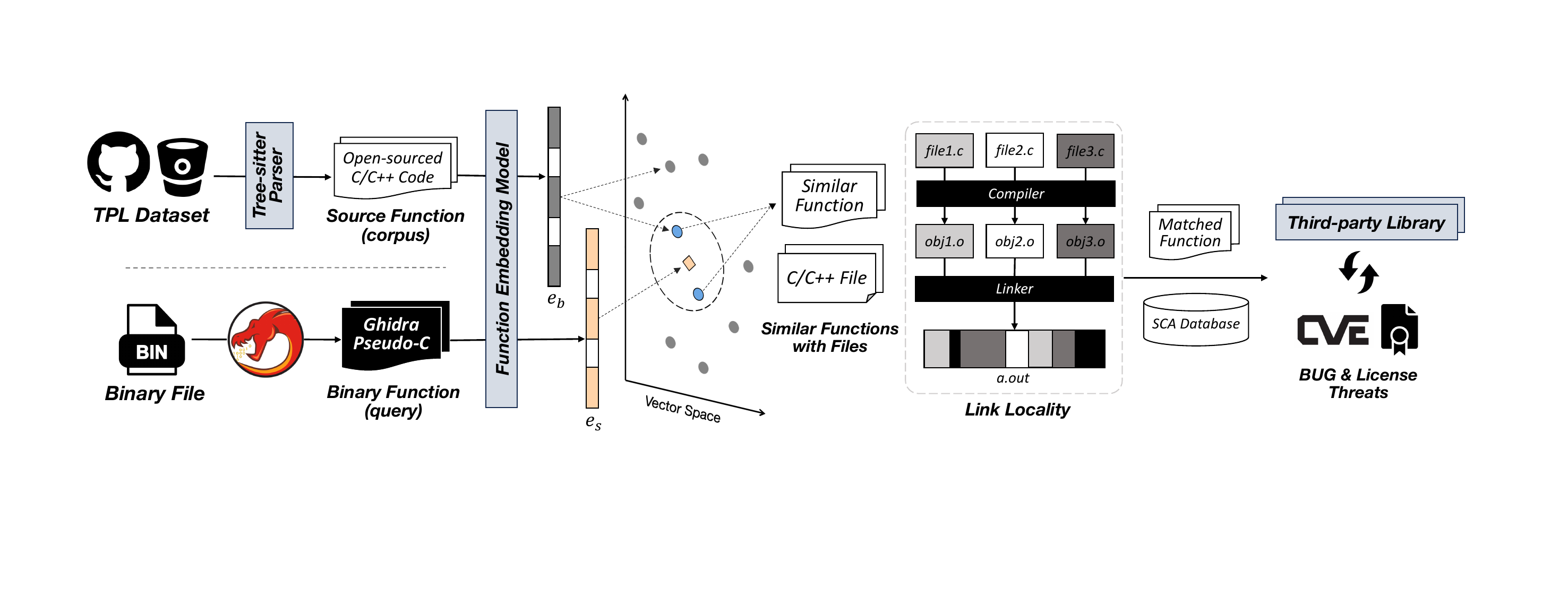}
    \caption{The workflow of \binaryai{}}
    \label{fig:binaryai}
\end{figure*}

\subsection{Motivation}
In this section, we intend to discuss the respective limitations of binary-to-binary and binary-to-source SCA to motivate our approach. 
Notably, binary-to-binary SCA can be compromised by the poor scalability of the TPL dataset. In particular, due to the intricacies associated with automatic compilation, only a limited subset of source packages maintained by package managers can be compiled automatically into multiple versions of binary files and incorporated into the SCA database. Extensive open-source C/C++ projects, such as GitHub repositories, can hardly be included in the TPL dataset, which is hindered by the substantial overhead of manual compilation. For instance, ModX~\cite{yang2022modx}, the state-of-the-art technique, selects 100 most frequently reused TPLs from a total of 795 maintained by Nix~\cite{nix} to build the binaries as the database. However, there are around 10K TPLs ($\sim$100X compared with ModX) in the existing largest dataset for binary-to-source SCA~\cite{tplite}. Note that the limited scale of the TPL dataset can significantly inhibit the practicality of SCA due to the likelihood that the contained TPLs and the corresponding vulnerabilities cannot be identified. Therefore, we select binary-to-source SCA as the primary subject of our investigation.

Subsequently, we deliberate the constraints of binary-to-source SCA. Existing binary-to-source SCA tools leverage basic syntactic features, such as string literals, to establish a correspondence between binary code and source code of TPLs, which may not well generalize to all the scenarios. Firstly, these basic features tend to exhibit a significant degree of redundancy in the large-scale TPL dataset. 
For instance, the string \textit{``407 Proxy Authentication Required''}, which indicates a common HTTP error, duplicates across more than 50 TPLs within our collected dataset. The presence of redundant syntactic features decreases their uniqueness and effectiveness, incurring inevitable false positives to decrease the precision of SCA. Furthermore, it is commonly observed that few or even no common syntactic features exist between reused TPLs and target binary files, especially the binary files which are stripped of distinctive features such as string literals and exported function names~\cite{yuan2019b2sfinder}. Meanwhile, existing techniques for extracting strings from C/C++ source code are not inherently robust, e.g., missing strings generated by concatenating macro-defined and constant strings to mismatch the string literals extracted from the binary files in the corresponding TPLs, such that the recall of binary-to-source SCA can also be compromised. Therefore, it is essential to employ fine-grained features (e.g. function-level features) 
in binary-to-source SCA such that high-level semantic information can be processed to mitigate the issue of redundancy and unreliability with basic features.

In this paper, considering the substantial disparities between binary and source functions introduced by compilation, we attempt to enhance binary-to-source SCA by adopting a transformer-based model to produce function-level embeddings and conducting binary source code matching accordingly. 

\section{Approach}


We propose \binaryai{}, a binary-to-source SCA technique with intelligent binary source code matching. Figure~\ref{fig:binaryai} presents the workflow of \binaryai{}, which consists of four phases: \textit{\approachfeature{}} (Section~\ref{sec:feature}), \textit{\approachembedding{}} (Section~\ref{sec:embedding}), \textit{\approachlocality{}} (Section~\ref{sec:locality}) and \textit{\approachsca{}} (Section~\ref{sec:sca}). Specifically, \binaryai{} is initialized by extracting C/C++ source-code functions 
from extensive repositories in the TPL dataset and C-like pseudo-code functions from the target binary file via decompilation (marked as \ding{182}). Accordingly, \binaryai{} adopts the large language model to generate the embeddings for source and binary functions. Note that binary source code matching in \binaryai{} is not an end-to-end process 
but is divided into two distinct stages. Specifically, \binaryai{} first trains a transformer-based model to learn the token-based syntactic features and retrieves \textit{top-k} most similar source functions from the corpus for each query binary function (\ding{183}). Next, \binaryai{} utilizes additional structured representations (e.g., \linklocality{}) to capture semantic features and match the exact source function from the \textit{top-k} candidates (\ding{184}). Eventually, \binaryai{} identifies the reused TPL components when the corresponding ratio of common source functions exceeds a pre-defined threshold with the target binary file (\ding{185}). 


\subsection{\ApproachFeature{}}
\label{sec:feature}

We extract functions with other meta information from both the TPL dataset and the target binary file respectively, in preparation for subsequent phases of \binaryai{}. Specifically, we characterize the features in terms of source function and binary function. 

\parabf{Source Function.} All the open-source projects in the TPL dataset employ \codeIn{git} as their version control system.  For each project, we collect the C/C++ source files across all the versions (i.e., git tags) and distinguish each file with the hash value of its content. Then we apply tree-sitter~\cite{treesitter}, an open-source source code parser, to construct the file's abstract syntax tree (AST) with its built-in C/C++ language parsers and extract all the unique source functions. Meanwhile, we maintain two inverted indexes to store the correspondences of extracted source functions into the SCA database, where one index maps each source function to all the files containing it, and the other maps each source function to all the  TPLs containing it.


\parabf{Binary Function.} We leverage Ghidra~\cite{ghidra}, an open-source reverse engineering framework developed by National Security Agency (NSA), to analyze the binary file, which involves disassembling the binary code and identifying functions, data structures, and other relevant information. Subsequently, Ghidra performs decompilation to generate the C-like pseudo code representation of the functions (i.e., binary functions). Additionally, we leverage Ghidra to extract the relative virtual address (denoted as \textit{bin\_rva}) as the ordinal number denoting the \linklocality{} in the binary file along with the function call graph as the inter-function communication. Note that we design \binaryai{} with the assumption that the input binary file has been stripped, i.e., all the debugging and and symbol information are eliminated, which is common in real-world scenarios~\cite{yuan2019b2sfinder}.  

\subsection{\ApproachEmbedding{}}
\label{sec:embedding}


The core insight of BinaryAI is to perform function-level binary source code matching based on function embeddings. In particular, our objective is to train a model that learns meaningful vector representations for both binary and source functions in a single vector space, where similar binary-to-source function pairs are expected to stay close while dissimilar ones are far apart. In this way, their similarity can be calculated using their corresponding embeddings. Typical code representation learning allows only one single code format of the matched objects, i.e., either source-to-source~\cite{sajnani2016sourcerercc, lopes2017dejavu, woo2021centris, fang2020functional, liu2023learning} or binary-to-binary~\cite{wang2022jtrans, Luo2023VulHawkCV, kim2022improving, xu2023improving, liu2018adiff} code matching. However, for binary source code matching, C/C++ language features (e.g., function inlining~\cite{jia20231}) and compiler optimization (e.g., code motion~\cite{knoop1994optimal}) can lead to substantial differences between binary code and source code, and such disparity can be rather challenging when designing \binaryai{}. To fill this gap, an ideal model needs to accurately capture subtle syntactic features to generate code embeddings for measuring similarity. Notably, existing large language models are extremely effective at learning the syntax of natural language~\cite{rothe2021simple, sun2022multilingual}, and this ability extends to code languages as well~\cite{feng2020codebert, guo21graphcodebert, ahmad2021plbart}. In particular, a model trained in multiple programming languages can potentially identify similar token-based features across different code formats~\cite{zeng2022extensive}. This can help detect code clones, even when the code has been translated into different languages. 
To this end, we use an existing large language model as the base model and further pre-train the model with a corpus of labeled binary source function pairs to build our model. Specifically, we apply a contrastive learning approach in a supervised manner to train the model acting as the function encoder to generate embeddings. This allows us to learn the code representations for both binary and source functions that minimize the distance between similar positive samples while distancing dissimilar negative samples. In this paper, we adopt \basemodel{}~\cite{biderman2023pythia}, which is widely adopted by the research community, as the base model to train our model\footnote{The base model for \binaryai{} has evolved through multiple iterations. Previously, we have also employed OPT~\cite{zhang2022opt} and BLOOM~\cite{workshop2022bloom} as the based models. By the time of publication, we use \basemodel{} because it has delivered optimal performance after training.}. As mentioned before, we initialize the original model with 410M parameters from the suite of \basemodel{} (i.e., \codeIn{pythia-410m}~\cite{pythia-hf}) and then further perform pre-training using contrastive learning.


\begin{figure}[htbp]
    \centering
    \includegraphics[width=0.98\columnwidth]{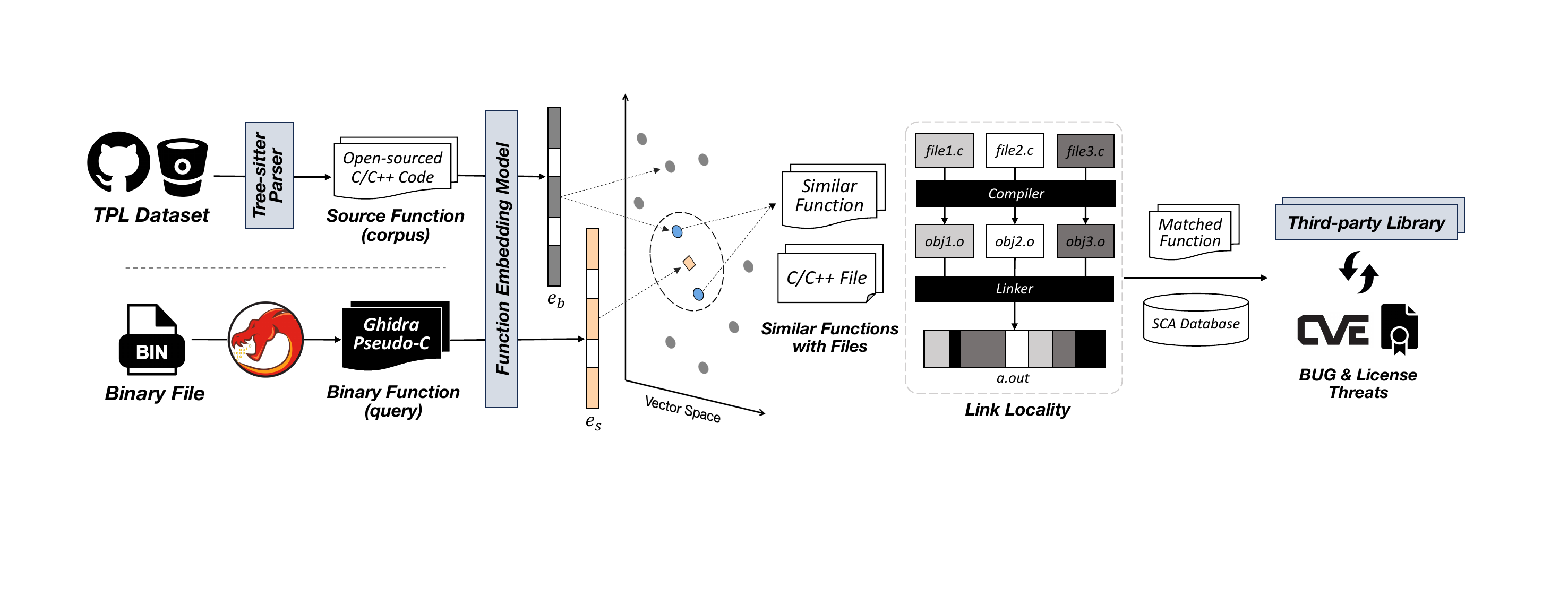}
    \caption{CLIP over binary and source function pairs}
    \label{fig:clip}
\end{figure}


Note that as one of the key ingredients in contrastive learning, enlarging in-batch negatives can effectively help the model to learn more discriminative representations as it needs to distinguish between a larger number of positive and negative samples in each batch~\cite{jain2021contra}, leading to better representation learning and improved performance on downstream tasks (i.e., binary SCA). To this end, we leverage the loss function of CLIP (Contrastive Language-Image Pre-training)~\cite{radford2021learning}, which is originally designed to align the representations of images and text captions, as our contrastive training objective. Figure~\ref{fig:clip} presents the training process based on CLIP contrastive learning method. We first perform tokenization that converts the functions into a sequence of tokens. Subsequently, we pass the tokenized input through the \basemodel{} model to obtain the function embeddings by extracting the output of the last hidden layer, where we denote $(e_i^{b}, e_i^{s})$ as the binary and source function embeddings of the $i_{th}$ positive sample. Accordingly, $(e_i^{b}, e_j^{b})$ represents one negative sample if $i$ is not equal to $j$. For the process of contrastive training, one batch consists of $N$ binary-to-source pairs and CLIP calculates the cosine similarity matrix between all the possible pairs. The training objective is to maximize the similarity between $N$ positive samples while minimizing the similarity for the rest $N*(N-1)$ negative samples via a symmetric cross-entropy loss over the matrix~\cite{weng2021contrastive}. Equation~\ref{eq:clip-bin} presents the binary-to-source loss function $L_{bin}$ and the source-to-binary loss function $L_{src}$ where $\tau$ is a learnable parameter to scale the logits. Note that the two loss functions are differed by swapping binary and source function embeddings when computing the similarity. 
Therefore, the overall loss function $L_{CLIP}$ is the average value of $L_{bin}$ and $L_{src}$ denoted as Equation~\ref{eq:clip}. Moreover, we extend the Momentum Contrast (MoCo) methodology~\cite{he2020momentum} to our contrastive pre-training, which further increases the number of negative samples and enables more effective contrastive learning by building dynamic dictionaries for CLIP.


\begin{equation}
L_{bin(src)} = -\frac{1}{N}\sum_{i=1}^{N}\log{\frac{\exp{(\text{sim}(e_i^{b(s)}, e_i^{s(b)})} / \tau)}{\sum_{j=1}^{N}\exp{(\text{sim}(e_i^{b(s)}, e_j^{s(b)}) / \tau)}}}
\label{eq:clip-bin}
\end{equation}

\begin{equation}
L_{CLIP} = (L_{bin} + L_{src}) / 2
\label{eq:clip}
\end{equation}

\begin{figure}[htbp]
    \centering
    \includegraphics[width=1.0\columnwidth]{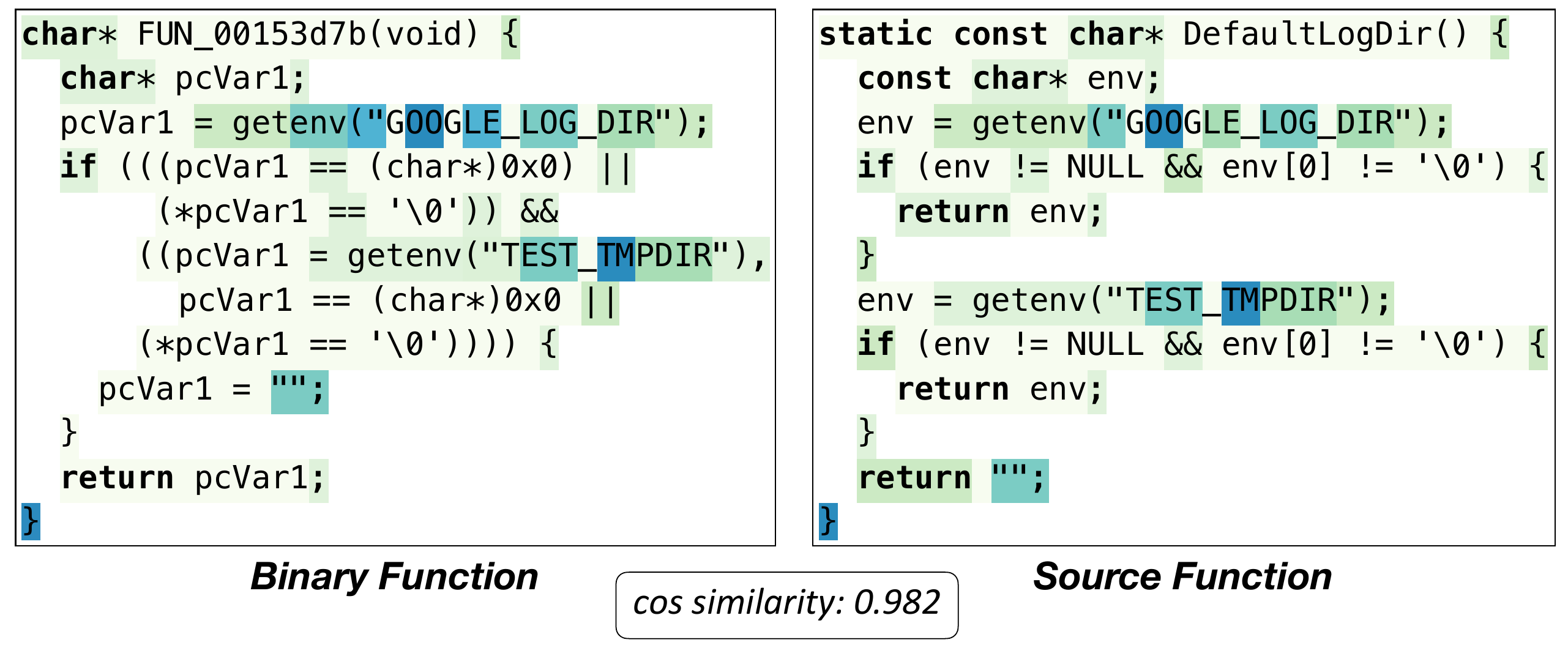}
    \caption{Retrieved binary and source function pair}
    \label{fig:positive-pair}
\end{figure}

We deploy the trained model in \binaryai{} by initially deriving function embeddings offline for all the source functions in the SCA database (with a total of \srcfuncnum{} unique functions from \tplnum{} TPLs) and store the source function embeddings to the vector database as corpus. Then for the online binary-to-source SCA task, we extract binary functions from the target binary file and perform real-time derivation of binary function embeddings. These derived embeddings serve as queries to retrieve similar source functions from the corpus for a given binary function.  Figure~\ref{fig:positive-pair} presents the retrieved \textit{top-1} most similar source function with the query of a binary function. This is actually a positive sample and has a similarity of 0.982. 
Eventually, we apply the relative virtual address (denoted as \textit{bin\_rva}) as the identifier for binary functions and obtain their \textit{top-k} most similar source functions as the output of \approachembedding{}. Note that we attach all the source files containing the corresponding functions (i.e., \textit{src\_func} $\Rightarrow$ \textit{src\_files}) 
by accessing the inverted index described in Section~\ref{sec:feature}.

\subsection{\ApproachLocality{}}
\label{sec:locality}

Ideally, we can directly select the source function with the highest similarity to the binary function in terms of function embeddings (i.e., \textit{top-1 of \approachembedding{}}) as our matching result. Nevertheless, due to the subtle modifications in source functions across different versions, there is a significant presence of similar functions within the large-scale TPL source repositories. Consequently, relying solely on language model-generated function embeddings to capture token-based syntactic features is insufficient for accurately matching the source functions, since the retrieved \textit{top-k} source functions can be quite similar. To tackle this issue, we attempt to leverage \linklocality{}~\cite{lfa} (i.e., relative virtual address as described in Section~\ref{sec:feature}) and \fcg{} as supplementary inter-function communication representing structured semantic features, which can help further identify the positive sample from the \textit{top-k} similar source functions in the second phase of binary source code matching. In this section, we present the fundamental rationale and the workflow of \approachlocality{}.

For the conventional C/C++ toolchain used to build binary files, the source code files (\textit{file.c}) are initially compiled into object files (\textit{obj.o}) by the compiler. Subsequently, the linker resolves symbol references between object files and combines them to produce the binary file, where the code sections of each object file are merged. By analyzing the process of compilation, we can derive several basic findings. 1) All the source functions in the same source file are compiled into a single object file although their relative locality to the source file might change. 2) The object files are continuously linked into the binary file, and all the functions (i.e., binary functions in the machine code format) within the code section of the object file preserve their relative locality. 3) Due to C/C++ template functions and conditional compilation, one source function in a source file can correspond to multiple binary functions in the object files (i.e., ``1-to-n'' mapping from source to binary functions). Inspired by these findings, we can further derive that the \linklocalities{} of the binary functions compiled from the same source file are rendered continuous in the binary file. Correspondingly, given the address space of the binary file, we can perform reverse engineering by cutting intervals containing continuous binary functions to recover the boundaries of the object files~\cite{lfa} and further identify the corresponding source files compiled into the binary file, thus accurately matching the source functions. To this end, we extract the continuous function pairs by \linklocality{} for each source file as the function intervals that can be mapped back to the address space of the binary file. Note that we consider isolated function pairs in the file as invalid matches and eliminate them while selecting the continuous interval. Compared to other files, the files compiled into the binary file should have a longer continuous interval of functions. Therefore, we form the file selection as an interval covering problem within the address space of the binary file and further utilize \fcg{} to facilitate the binary source function matching within the selected files.

\SetKwFor{While}{while}{do}{}
\SetKwInput{KwInput}{Input}
\SetKwInput{KwResult}{Result}
\SetKw{KwReturn}{return}
\SetAlCapFnt{\footnotesize}
\SetKwComment{Comment}{$\triangleright$\ }{}
\SetCommentSty{mycommfont}
\DontPrintSemicolon
\begin{algorithm}[t]
\footnotesize
\setstretch{1.1}
\caption{\footnotesize \ApproachLocality{}}
\label{alg:link-locality}
    \KwInput{\textit{bin2src\_topk} \Comment*[r]{Retrieved topk similar source functions}}
    \KwResult{\textit{bin2src\_match} \Comment*[r]{Matched binary source function pairs}}
    
    \SetKwProg{Fn}{Function}{:}{}
    \Fn{MatchFuncPairs} {
        \textit{file2pairs, intervals, bin2src\_match} $\leftarrow$ $\emptyset$ \\
        \For{(\textit{bin\_rva}, \textit{similar\_funcs}) $\in$ \textit{bin2src\_topk}} {
            \For{(\textit{src\_func}, \textit{src\_files}) $\in$ \textit{similar\_funcs}} {
                \textit{file2pairs[file].add(bin\_rva $\Rightarrow$ src\_func)} \textbf{for} \textit{file} in \textit{src\_files}
            }
            \textit{bin2src\_match[bin\_rva]} = \textit{top1\_similar\_src\_func} \\
        }
        \For{(file, bin2src\_pairs) $\in$ file2pairs} {
            \textit{intervals.add(MaxFileInterval(bin2src\_pairs, file.func\_count)}
        }
        \textit{intervals.sort(key=lambda x : (x.start, -x.end, x.max\_hit))} \\
        \For{interval in intervals} {
            \If{interval.start > last\_select\_interval.end}{
                \For{(bin\_rva, src\_func) $\in$ RestrictByFCG(interval.func\_pairs)} {
                    \textit{bin2src\_match[bin\_rva]} = \textit{src\_func} \\
                }
                \textit{last\_select\_interval} $\leftarrow$  \textit{interval}
            }
        }
        \KwRet \textit{bin2src\_match}
    }
    \;
    \SetKwProg{Fn}{Function}{:}{}
    \Fn{MaxFileInterval(bin2src, file\_func\_count, i=0, j=0)} {
        \textit{address} $\leftarrow$ \textit{sort(bin2src.keys)} \\
        \textit{interval} $\leftarrow$ \textit{initInterval(max\_hit=0)}\\ 
        \While{\textit{j} < len(\textit{address})} {
            \textit{func\_slice} $\leftarrow$ \textit{\{ bin\_rva : bin2src[bin\_rva] for bin\_rva in address[i : j] \}} \\
            \textit{src\_hit} $\leftarrow$ \textit{len(func\_slice.values)} \\ 
            \uIf{src\_hit <= \textit{file\_func\_count}} {
                \If{src\_hit > interval.max\_hit} {
                   \textit{interval.max\_hit} $\leftarrow$ \textit{src\_hit} \\
                   \textit{interval.start, interval.end} $\leftarrow$ \textit{i, j} \\
                   \textit{interval.func\_pairs} $\leftarrow$ \textit{func\_slice} \\
                }            
                \textit{j++}
            }
            \textbf{else} \textit{i++}
        }
        \KwRet \textit{interval}
    }
\end{algorithm}

Algorithm~\ref{alg:link-locality} presents the overall workflow of \approachlocality{}. 
First, we obtain all the included files of each source function from retrieved \textit{top-k} candidates and build the index \textit{file2pairs} mapping each source file to all its retrieved binary source function pairs (lines 3-5). 
Meanwhile, we initialize the matching result with the \textit{top-1} most similar source function (line 6). Next, we extract the continuous function pairs for each source file (lines 7-8). Specifically, we sort the function pairs by \textit{bin\_rva} (i.e., \linklocality{}) and leverage a sliding window with two separate pointers ($i \& j$, both are initialized to 0) to slice the file and generate the corresponding function interval (lines 17-30) that simultaneously satisfies two conditions. 1) The number of the source functions within the interval does not exceed the total number of functions in the file (line 23). 2) The sliced interval has the maximum number of function pairs (lines 24-25). Furthermore, we map the continuous function intervals extracted from each source file back to the address space of the binary file based on \textit{bin\_rva}. As mentioned before, we attempt to select longer intervals to cover as many functions within the binary file as possible. To this end, we transform the file selection into an interval covering problem and tackle the problem greedily. Specifically, we sort the intervals according to a particular set of priorities (line 9), i.e., interval with lower start point (\textit{x.start}), higher end point (\textit{-x.end}), and more contained binary source function pairs (\textit{x.max\_hit}), which allows us to prioritize longer intervals that can also cover more binary functions compared with other equally long intervals. Before selecting the interval corresponding to the source file, we ensure that its start point should be higher than the end point of the previously selected interval to avoid potential overlaps (lines 10–11).

\begin{figure}[htbp]
    \centering
    \includegraphics[width=1.0\columnwidth]{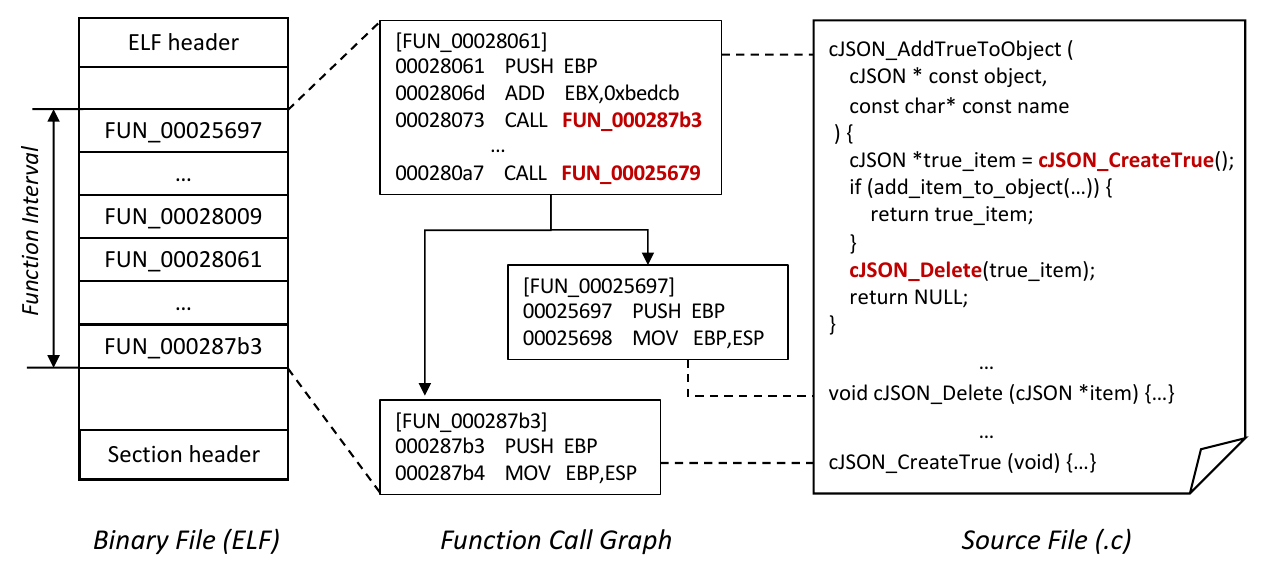}
    \caption{Function matching between source and binary files}
    \label{fig:link-locality}
\end{figure}

As in Finding 3, while there can be ``1-to-n'' correspondence from source to binary functions in a single source file, the retrieved function pairs in the file tend to be more complicated and elusive (e.g., ``n-to-n'' mappings caused by similar source functions). Correspondingly, incorrect function matching might still occur even when we identify the correct source file. To alleviate the issue, we leverage \fcg{} to restrict the function pairs for each selected file before generating the matching results (Algorithm~\ref{alg:link-locality}, line 12). Suppose there are two function pairs ($bin_1$, $src_1$) and ($bin_2$, $src_2$) in the source file. If there are function calls between both the binary and source functions (e.g., $bin_1$ calls $bin_2$ and $src_1$ calls $src_2$), these two function pairs can be considered correct. In this case, we can filter out the mapping from these binary functions to other source functions, e.g., ($bin_1$, $src_3$). Figure~\ref{fig:link-locality} presents an example of function matching between source and binary files with the longest function interval $[25697, 287b3]$. We can observe from the \fcg{} that both the binary function \codeIn{FUN\_00028061} and source function \codeIn{cJSON\_AddTrueToObject} in the function pair have two callees, which are also matched in the same file. Therefore, we can derive three correct function matches via \fcg{}. Eventually, we assign all remaining function pairs in the selected files to update the matching results of binary-to-source functions (line 13) as the output of \approachlocality{}. 

\subsection{\ApproachSCA{}}
\label{sec:sca}

\SetKwFor{While}{while}{do}{}
\SetKwInput{KwInput}{Input}
\SetKwInput{KwResult}{Result}
\SetKw{KwReturn}{return}
\SetAlCapFnt{\footnotesize}
\SetKwComment{Comment}{$\triangleright$\ }{}
\SetCommentSty{mycommfont}
\DontPrintSemicolon
\begin{algorithm}[t]
\footnotesize
\setstretch{1.05}
\caption{\footnotesize \ApproachSCA{}}
\label{alg:sca}
    \KwInput{\textit{bin2src\_match, tpl\_dependency}}
    \KwResult{\textit{components}}
    


    \SetKwProg{Fn}{Function}{:}{}
    \Fn{DetectComponents} {
        \textit{tpl2func\_match, components} $\leftarrow$ $\emptyset$ \\
        \For{(bin\_rva, src\_func) $\in$ \textit{bin2src\_match}} {
            \textit{src\_tpls} $\leftarrow$ retrieved TPLs containing \textit{src\_func} in SCA database \\
            \textit{filtered\_tpls} $\leftarrow$ \textit{FilterByDependency(src\_tpls, tpl\_dependency)} \\    
            \For{tpl $\in$ filtered\_tpls} {
                \textit{tpl2func\_match[tpl].add(bin\_rva)}
            }
        }
        \For{(tpl, matched\_funcs) $\in$ \textit{tpl2func\_match}} {
            \If{\textit{len(matched\_funcs) / tpl.total\_func\_count} > $\theta$} {
                \textit{components.add(tpl)}
            }
        }
        \KwRet \textit{components}
    }
    \;
    \SetKwProg{Fn}{Function}{:}{}
    \Fn{FilterByDependency(src\_tpls, tpl\_dependency)} {
        \textit{filtered\_tpls} $\leftarrow$ \textit{src\_tpls} \\
        \For{tpl $\in$ src\_tpls} {
            \textit{reused\_tpls} $\leftarrow$ \textit{tpl\_dependency[tpl]} \\
            \If{reused\_tpls and src\_tpls have intersection} {
                \textit{filtered\_tpls.remove(tpl)}
            }
        }
        \KwRet \textit{filtered\_tpls}
    }
\end{algorithm}

\binaryai{} acquires the matched source functions and further performs TPL detection (i.e., SCA task) for the target binary file as presented in Algorithm~\ref{alg:sca}. The baseline technique is to retain all the included TPLs of each matched source function by referring to the SCA database (lines 3-4), which contains the inverted index of the correspondence from the source functions to TPLs as described in Section~\ref{sec:feature}. However, noticing that in general source functions are extensively cloned across various TPLs~\cite{tang2022towards, lopes2017dejavu, sajnani2016sourcerercc, Woo2022MOVERYAP} in the large-scale dataset, such internal code clones~\cite{duan2017identifying} can lead to inevitable false positives if we retain all their included TPLs. For instance, assuming a binary file only contains the TPL \textit{zlib} as the component, other TPLs reusing \textit{zlib} are also identified as the components owing to the common functions cloned from \textit{zlib}. To alleviate the issue, we follow previous works~\cite{woo2021centris, ossfp, tplite} to filter TPLs based on the TPL dependency which exhibits the reuse relations across TPLs and only retain the reused ones. Specifically, we leverage \tplite{}~\cite{tplite}, the state-of-the-art technique based on function birth time (i.e., the earliest release time) and hierarchical path information, to generate the TPL dependency in advance, which works as the additional input of SCA to help identify the reused TPLs.

Algorithm~\ref{alg:sca} presents the workflow of TPL detection in \binaryai{}. First, we extract all the included TPLs for each matched source function from the SCA database (lines 3-4). Subsequently, we filter TPLs based on the TPL dependency and count the matched functions for all the retained TPLs (lines 5-7). Specifically, we filter the TPLs whose reused ones are also included with the matched source function (lines 14-18), which should indicate the internal code clones between TPLs. For instance, the matched source function \codeIn{deflateInit} is both included in TPLs \textit{zlib}~\cite{zlib} (deflate.c) and \textit{llvm}~\cite{llvm} (llvm/runtime/zlib/deflate.c). The TPL dependency indicates \textit{llvm} reuses \textit{zlib}, and we thus only select \textit{zlib} by filtering out \textit{llvm}. Eventually, we calculate the ratio of matched functions to the total number of source functions for each selected TPL, indicating the similarity between the binary file and the source code repository. If the ratio exceeds a pre-defined threshold $\theta$, \binaryai{} identifies the corresponding TPL as the contained component in the target binary file (lines 8-10). Meanwhile, \binaryai{} detects whether security threats are introduced by these components by retrieving the official vulnerability repository, e.g., the NVD database~\cite{21841}.

\section{Evaluation}

In the evaluation, we attempt to investigate the performance of \binaryai{} by answering the following research questions:

\begin{itemize}[leftmargin=*]
\item \textit{\textbf{RQ1:}} How effective is the embedding model in measuring the similarity between the binary and source functions?

\item \textit{\textbf{RQ2:}} How does \binaryai{} perform in terms of binary source code matching with the two separate phases?

\item \textit{\textbf{RQ3:}} What is the accuracy of \binaryai{} in detecting TPLs in binary files compared to state-of-the-art techniques?

\end{itemize}

\subsection{Dataset}

To extensively evaluate the performance of \binaryai{} in terms of different mechanisms, we first construct three datasets following existing works~\cite{tang2022libdb, yang2022modx, yuan2019b2sfinder, duan2017identifying, woo2021centris} for model training and the evaluation of the downstream binary-to-source SCA task.

\subsubsection{Training Dataset} 
\label{sec:train-data}

To obtain a large number of matched binary-to-source function pairs as positive samples for training the model, we construct the automatic compilation pipeline based on official ArchLinux packages~\cite{Arch_linux} and Arch User Repository (AUR)~\cite{Arch_User_Repository} following the insight from BinaryCorp in jTrans~\cite{wang2022jtrans}. Specifically, we apply the command \codeIn{makepkg} to compile all the ArchLinux packages and AUR automatically. Meanwhile, we hook the compiler to generate the debugging information in the format of DWARF~\cite{DWARF}. On one hand, we decompile the output binary file with Ghidra~\cite{ghidra} to acquire the mapping from the virtual address to the binary function. On the other hand, we parse the DWARF debugging information and extract the mapping from the virtual address to the source file with line number. We further leverage tree-sitter~\cite{treesitter} to slice out the corresponding source function in the file. By merging the mapping from both sides and filtering out mismatched functions due to runtime errors, we obtain around \totalpair{} matched function pairs with an average of about 500 tokens per function as the training set.

\subsubsection{TPL Dataset \& Corpus}
\label{sec:tpl-data}
Following previous works~\cite{woo2021centris,tang2022towards,ossfp}, we collect a large number of C/C++ open-source projects by crawling from GitHub repositories and source packages of the GNU/Linux community, and we obtain the dataset consisting of \tplnum{} TPLs, which is adequate for the SCA task~\cite{woo2021centris}.
Next, we extract \srcfuncnum{} unique source functions\footnote{As of the time of publication, the magnitude is around 56M. Note that we deploy this module in industry, enabling continuous supplementation of new TPLs and source functions to improve the practicality of \binaryai{}.} and derive the corresponding function embeddings based on the trained model 
which are stored persistently in the FAISS~\cite{johnson2019billion} database as the corpus. To our best knowledge, this corpus is the largest in the domain of binary source code matching, where the state-of-the-art technique \codecmr{}~\cite{yu2020codecmr} retrieves close embeddings within the corpus of 10,000 functions. 
A larger corpus is more practical as it includes more source functions that are similar to each other. This significantly increases the difficulty of embedding-based function retrieval and further validates the generality of our mechanism.

\subsubsection{SCA Test Set}
\label{sec:sca-data}
To evaluate the performance of TPL detection for \binaryai{}, we construct our binary SCA test set compiled by \scaproduct{} open-source software projects and obtain \scabin{} binary files as the test cases, along with manually labeled components. Specifically, we collect highly prominent projects with more than 1K stars from GitHub. Furthermore, we select projects with over 10 sub-modules indicating that their compiled binaries are more likely to have multiple components, facilitating the evaluation of SCA. Next, we compile the source code of \scaproduct{} projects into \scabin{} stripped binary files across multiple architectures and compiler configurations. Meanwhile, we follow previous works~\cite{woo2021centris,tang2020libdx,tang2022libdb} to manually label the reused TPLs as the components by rigorously analyzing all file paths, included header files, and other meta-information from SBOM files (e.g., CMakeLists), README, copyright, and license. As a result, a total of \scalabel{} components are labeled as the ground-truth SCA results, forming the largest dataset in binary-to-source SCA. 

To further investigate the accuracy of \approachlocality{} and its contribution to binary source function matching, we need to label the ground-truth correspondence between binary and source functions at a fine granularity within real-world binary files. Given the high expense of manual analysis, we label the binary-to-source function mappings for \scare{} (10\%) most commonly used binary files out of \scabin{} binary files. In particular, we perform reverse engineering manually to determine which function within the source files is compiled to the binary function in the object files. As a result, we obtain \scarefunc{} matched function pairs within these \scare{} binary files.

\begin{table*}[htbp]
  \centering
  \caption{Result of retrieving similar source functions}
  \label{tab:smart-func}
  \setlength\tabcolsep{3pt}
  \renewcommand\arraystretch{1.05}
  \begin{adjustbox}{width=0.98\textwidth}
  \begin{tabular}{lcccccc|ccccc}
  \toprule[1.2pt]
  \multirow{2}{*}{\textbf{Model}} & \multirow{2}{*}{\textbf{Objective}} & \multicolumn{5}{c}{\textbf{Validation Set of Model (query=32,296)}} & \multicolumn{5}{c}{\textbf{Binary SCA Test Set (query=\scarefunc{})}} \\
  \cmidrule{3-12}
    & & MRR & Count/Recall@1 &  Count/Recall@10 & Count/Recall@50 & Count/Recall@100 & MRR & Count/Recall@1 &  Count/Recall@10 & Count/Recall@50 & Count/Recall@100 \\
  \midrule
  \binpro{}   & N/A    & \multicolumn{1}{|c}{0.0027} &   771 /  2.39 &  1,165 /  3.61 &  1,593 /  4.93 &  1,845 /  5.71 & 0.0036 &  612  /  2.60 &    944 /  4.01 &  1,262 / 5.36  &   1,507 /  6.40 \\
  \bsfinder{} & N/A    & \multicolumn{1}{|c}{0.0042} &   945 /  2.93 &  1,717 /  5.32 &  2,108 /  6.53 &  2,436 /  7.54 & 0.0048 &  864  /  3.67 &  1,305 /  5.55 &  1,740 / 7.40  &   2,082 /  8.85 \\ \midrule
  \codecmr{}  & Triplet  & \multicolumn{1}{|c}{0.1431} & 3,195	/  9.89 &  6,543 / 20.26 &  7,827 / 24.24 &  8,347 / 25.85 & 0.2232 & 2,805	/ 11.92 &  7,873 / 33.46 &  9,875 / 41.97 &  1,0561 / 44.89 \\
  \codecmr{}  & CLIP     & \multicolumn{1}{|c}{0.2319} & 5,456	/ 16.89 & 10,589 / 32.79 & 12,256 / 37.95 & 12,801 / 39.64 & 0.2820 & 3,638	/ 15.46 &  9,889 / 42.03 & 12,510 / 53.17 &  13,319	/	56.61 \\
  \binaryai{} & Triplet  & \multicolumn{1}{|c}{0.2774} & 6,552 / 20.29 & 12,627 / 39.10 & 14,009 / 43.38 & 14,460 / 44.77 & 0.3539 & 4,692 / 19.94 & 12,113 / 51.48 & 14,650 / 62.26 &  15,395	/ 65.43 \\
  \midrule
  \textbf{\binaryai{}} & \textbf{CLIP} & \multicolumn{1}{|c}{\textbf{0.3006}} & \textbf{7,235 / 22.40} & \textbf{13,465 / 41.69} & \textbf{14,682 / 45.46} & \textbf{15,020 / 46.51} & \textbf{0.3958} & \textbf{5,348 / 22.73} & \textbf{13,493 / 57.35} & \textbf{15,873 / 67.46} &  \textbf{16,576 / 70.45} \\
  \bottomrule[1.2pt]
  \end{tabular}
  \end{adjustbox}
\end{table*}

\subsection{Experiment Setup}
\label{sec:setup}
To evaluate the effectiveness of \approachembedding{}, we include \codecmr{}~\cite{yu2020codecmr}, the state-of-the-art binary source function matching model, for performance comparison with the model of \binaryai{}. Note that \codecmr{}  utilizes separate function encoders (DPCNN for source function and GNN for binary function) and triplet loss as the contrastive learning objective. To ensure a fair comparison, we adopt the same training set described in Section~\ref{sec:train-data}. For the training process, we follow the training setup in the original paper for \codecmr{}. As for \binaryai{}, the maximum length of the embedding model is 2048, the training epoch is 196, the batch size is 512, and the learning ratio is 0.001.
Furthermore, we follow \codecmr{} to include \binpro{}~\cite{miyani2017binpro} and \bsfinder~\cite{yuan2019b2sfinder} as traditional techniques in comparison with the neural network-based techniques for binary source code matching. Both \binpro{} and \bsfinder{} match code with basic syntactic features (e.g., string and integer constants), and we use Hungarian algorithm~\cite{munkres1957algorithms} based on the weights in their original papers to match source functions from the corpus. Note that we adopt \testfunc{} and \scarefunc{} function pairs respectively from the validation set of the model and the \scare{} manually labeled SCA test cases as the query sets to validate whether the model can generalize to different datasets.

Given binary functions as queries, we adopt multiple metrics to evaluate the performance of retrieving similar source functions from the corpus. Specifically, we adopt \textit{MRR} (Mean Reciprocal Rank) computed by averaging the reciprocal ranks across all queries as denoted in Equation~\ref{eq:mrr}. To verify the upper bound of model capability and the effectiveness of \approachlocality{}, we also adopt the count of positive samples that can be detected within retrieved \textit{top-k} similar functions and the corresponding recall by dividing it to the total number of queries (denoted as \textit{Count/Recall@k}).

\begin{equation}
MRR = \frac{1}{|Q|}\sum_{i=1}^{|Q|}\frac{1}{rank_i}
\label{eq:mrr}
\end{equation}

For \approachlocality{}, we evaluate the accuracy by identifying positive binary source function pairs, as well as its contribution to refining the \textit{top-1} result retrieved by the model, where we adopt the \scare{} binary files with manually labeled function mappings as the ground truth. Eventually, we utilize all the \scabin{} binary files with \scalabel{} labeled reused TPLs for the SCA task. We compare \binaryai{} with the existing binary-to-source SCA tools including two academic tools: \osspolice{}~\cite{duan2017identifying} and \bsfinder{}~\cite{yuan2019b2sfinder}, and two well-established commercial products: \scantist{} and \bdba{}\footnote{The specific commercial tools are not disclosed due to the constraints of confidentiality agreements and the proprietary nature of the software, thus the names of the commercial tools used in this paper have been anonymized.}. For a fair comparison, we utilize the same TPL dataset (\tplnum{} projects) to build their corresponding SCA database. 



    


Note that we adopt the same metrics for investigating these two tasks that include \textit{Precision} (i.e., the ratio of true positives to all the derived results), \textit{Recall} (i.e., the ratio of true positives to all the ground-truth data), and \textit{F1} score (i.e., the measure of accuracy by considering both precision and recall). Considering the trade-off between precision and recall, we set the threshold $\theta$ to 0.01 that achieves the maximum \textit{F1} score.

\subsection{Results and Analysis}

\subsubsection{RQ1: Effectiveness of Function Embedding}


We first compare \binaryai{} with \codecmr{} in terms of \approachembedding{}.
Table~\ref{tab:smart-func} demonstrates the evaluation results of retrieving source functions with two query sets as described in Section~\ref{sec:setup}.  We can observe that \binaryai{} outperforms \codecmr{} in both query sets in terms of \textit{MRR} (0.3006 vs. 0.1431 and 0.3958 vs. 0.2232). Moreover, \binaryai{} can retrieve more positive samples for all \textit{top-k} setups (i.e., \textit{Count/Recall@k}) compared with \codecmr{}. For instance, within the \textit{top-10} retrieved results with queries from the binary SCA test set, \binaryai{} detects 13,493 matched source functions for all the 23,529 queries with 57.35\% recall while \codecmr{} detects 7,873 with 33.46\% recall. By combining two query sets, \binaryai{} achieves 0.3407 \textit{MRR} in contrast to 0.1769 of \codecmr{}, indicating that the positive samples retrieved by \binaryai{} tend to have a higher average rank (around the 3rd, 1/0.34). Additionally, \binaryai{} effectively increases the \textit{recall@1} from 10.75\% to 22.54\% and \textit{recall@100} from 33.87\% to 56.60\% compared with \codecmr{}.

Note that \binaryai{} and \codecmr{} respectively employ a large language model (i.e., \basemodel{}) and a combination of DPCNN and GNN as the base model, along with CLIP loss and triplet loss as the training objective. To perform an ablation study on the model and training objective respectively, we further train two new models by reassembling the base models and training objectives from \binaryai{} and \codecmr{}. Table~\ref{tab:smart-func} demonstrates that \codecmr{} increases \textit{MRR} from 0.1431 to 0.2319 and \textit{recall@1} from 9.89\% to 16.89\% in the validation set by updating the loss function from triplet loss to CLIP. For \binaryai{}, modifying the training objective from CLIP to triplet loss degrades the performance (0.3006 vs. 0.2774 for \textit{MRR}), indicating the effectiveness of the loss function from CLIP. Then we investigate the impact of the base model, we can observe that even though \codecmr{} improves the performance by using CLIP, this effect is still inferior to \binaryai{} trained with triplet loss (0.2319 vs. 0.2774 for \textit{MRR}). Therefore, we can demonstrate the advantage of utilizing a large language model in the domain of binary source function matching.


\mybox{\textbf{Finding 1:} \binaryai{} can be more effective than \codecmr{} in terms of the \approachembedding{} with the usage of LLM and CLIP as the training objective.}

We further investigate the difference between neural network-based techniques (i.e., \binaryai{} and \codecmr{}) and the existing feature-matching-based techniques (i.e., \binpro{} and \bsfinder{}). We can observe that the performance of the traditional techniques is rather limited in retrieving matched source functions. Specifically, \textit{MRR} for \binpro{} and \bsfinder{} in both query sets is less than 0.005, indicating that the matched source function has a rank of over 200 on average for each query. Moreover, both \binpro{} and \bsfinder{} recall less than 10\% positive samples within \textit{top-100} and less than 5\% at \textit{top-1} with the two query sets. Next, we investigate the reason and find several factors leading to decreased performance. Firstly, many source functions in the corpus share similar basic features, making it challenging to distinguish them effectively. Secondly, some binary functions as queries lack meaningful basic features, further compromising the retrieved results.

\mybox{\textbf{Finding 2:} The existing feature-matching-based techniques incur limited performance in matching source functions from large-scale corpus, further indicating the effectiveness of \approachembedding{}.}

\subsubsection{RQ2: Accuracy of Binary Source Code Matching}

Previous findings indicate that \binaryai{} achieves distinct improvement in retrieving source functions from large-scale corpus compared with the state-of-the-art techniques. However, \binaryai{} is still limited to binary source code matching by directly applying \textit{recall@1} (22.73\% for \scarefunc{} queries from the binary SCA test set in Table~\ref{tab:smart-func}), which is insufficient for the downstream SCA task. In this RQ, we investigate the accuracy of \approachlocality{} along with its contribution to binary source code matching based on the SCA test set with \scare{} binary files. Table~\ref{tab:smart-re} presents the matching results with the input of retrieved \textit{top-10} similar source functions. Note that in addition to the results that match the ground truth (denoted as ``Exact Match''), we also follow previous works~\cite{kim2017vuddy,woo2021centris} to include the results that are identical to ground truth after normalization (denoted as ``Fuzzy Match'') since such results are applicable for other downstream tasks that do not require high accuracy, such as reverse engineering~\cite{david2020neural}. Overall, we can observe that the precision for the exact match is 81.63\% on average. More specifically, the precision exceeds 75\% in all binary files ranging from 75.19\% (\textit{turbobench}) to 94.92\% (\textit{hyriseSystem}). Such results illustrate that the accuracy of function matching based on \linklocality{} and \fcg{} is high and can generalize to all the binary files in the SCA test set. Moreover, the precision for the fuzzy match is 95.86\% on average and exceeds 90\% in all binary files ranging from 90.00\% (\textit{turbobench}) to 98.30\% (\textit{nano\_node}), i.e., a large amount of false positives can match the ground truth after normalization.

\mybox{\textbf{Finding 3:} Locality-driven matching can effectively identify the exact source function from top-k retrieved results and such results generalize to different binary files.}

\begin{table}[htbp]
  \centering
  \caption{Result of locality-driven matching (\textit{k}=10)}
  \label{tab:smart-re}
  \setlength\tabcolsep{5pt}
  \renewcommand\arraystretch{1.05}
  \begin{adjustbox}{width=0.98\columnwidth}
  \begin{tabular}{lccccc|ccc}
  \toprule[1.2pt]
  \multirow{2}{*}{\textbf{Binary}} &  \multirow{2}{*}{\textbf{\#Label}} & \multirow{2}{*}{\textbf{\binaryai{}}} & \multicolumn{3}{c}{\textbf{Exact Match}} & \multicolumn{3}{c}{\textbf{Fuzzy Match}} \\
  \cmidrule{4-9}
    &  &  & \#TP  & P (\%)  & R (\%) & \#TP  & P (\%) & R (\%) \\
  \midrule
  controlblock & \multicolumn{1}{|c}{185} & 107 & \multicolumn{1}{|c}{86} & 80.37 & 46.49 & 99 & 92.52 & 53.51 \\
  db\_bench & \multicolumn{1}{|c}{359} & 253 & \multicolumn{1}{|c}{209} & 82.61 & 58.22 & 239 & 94.47 & 66.57 \\
  dosbox\_core & \multicolumn{1}{|c}{2,804} & 2,042 & \multicolumn{1}{|c}{1,854} & 90.79 & 66.12 & 1,974 & 96.67 & 70.40 \\
  eth\_sc & \multicolumn{1}{|c}{267} & 232 & \multicolumn{1}{|c}{190} & 81.90 & 71.16 & 221 & 95.26 & 82.77 \\
  hyriseSystem & \multicolumn{1}{|c}{318} & 197 & \multicolumn{1}{|c}{187} & 94.92 & 58.81 & 193 & 97.97 & 60.69 \\
  kvrocks & \multicolumn{1}{|c}{2,240} & 1,452 & \multicolumn{1}{|c}{1,190} & 81.96 & 53.13 & 1,415 & 97.45 & 63.17 \\
  nano\_node & \multicolumn{1}{|c}{1,604} & 939 & \multicolumn{1}{|c}{752} & 80.09 & 46.88 & 923 & 98.30 & 57.54 \\
  pagespeed & \multicolumn{1}{|c}{6,430} & 3,442 & \multicolumn{1}{|c}{2,683} & 77.95 & 41.73 & 3,305 & 96.02 & 51.40 \\
  prometheus & \multicolumn{1}{|c}{204} & 157 & \multicolumn{1}{|c}{138} & 87.90 & 67.65 & 146 & 92.99 & 71.57 \\
  replay-sorcery & \multicolumn{1}{|c}{770} & 454 & \multicolumn{1}{|c}{367} & 80.84 & 47.66 & 437 & 96.26 & 56.75 \\
  st-device-sdk & \multicolumn{1}{|c}{801} & 582 & \multicolumn{1}{|c}{486} & 83.51 & 60.67 & 536 & 92.10 & 66.92 \\
  tendisplus & \multicolumn{1}{|c}{2,197} & 1,541 & \multicolumn{1}{|c}{1,265} & 82.09 & 57.58 & 1,498 & 97.21 & 68.18 \\
  tic80 & \multicolumn{1}{|c}{832} & 695 & \multicolumn{1}{|c}{573} & 82.45 & 68.87 & 668 & 96.12 & 80.29 \\
  turbobench & \multicolumn{1}{|c}{762} & 270 & \multicolumn{1}{|c}{203} & 75.19 & 26.64 & 243 & 90.00 & 31.89 \\
  yuzu-cmd & \multicolumn{1}{|c}{3,756} & 1,795 & \multicolumn{1}{|c}{1,374} & 76.55 & 36.58 & 1,675 & 93.31 & 44.60 \\
  \midrule
  \textbf{Total} & \multicolumn{1}{|c}{\textbf{23,529}} & \textbf{14,158} & \multicolumn{1}{|c}{\textbf{11,557}} & \textbf{81.63} & \textbf{49.12} & \textbf{13,572} & \textbf{95.86} & \textbf{57.68} \\
  \bottomrule[1.2pt]
  \end{tabular}
  \end{adjustbox}
\end{table}

\begin{figure}[htbp]
    \centering
    \begin{subfigure}[t]{0.49\columnwidth}
        \centering
        \includegraphics[width=\textwidth]{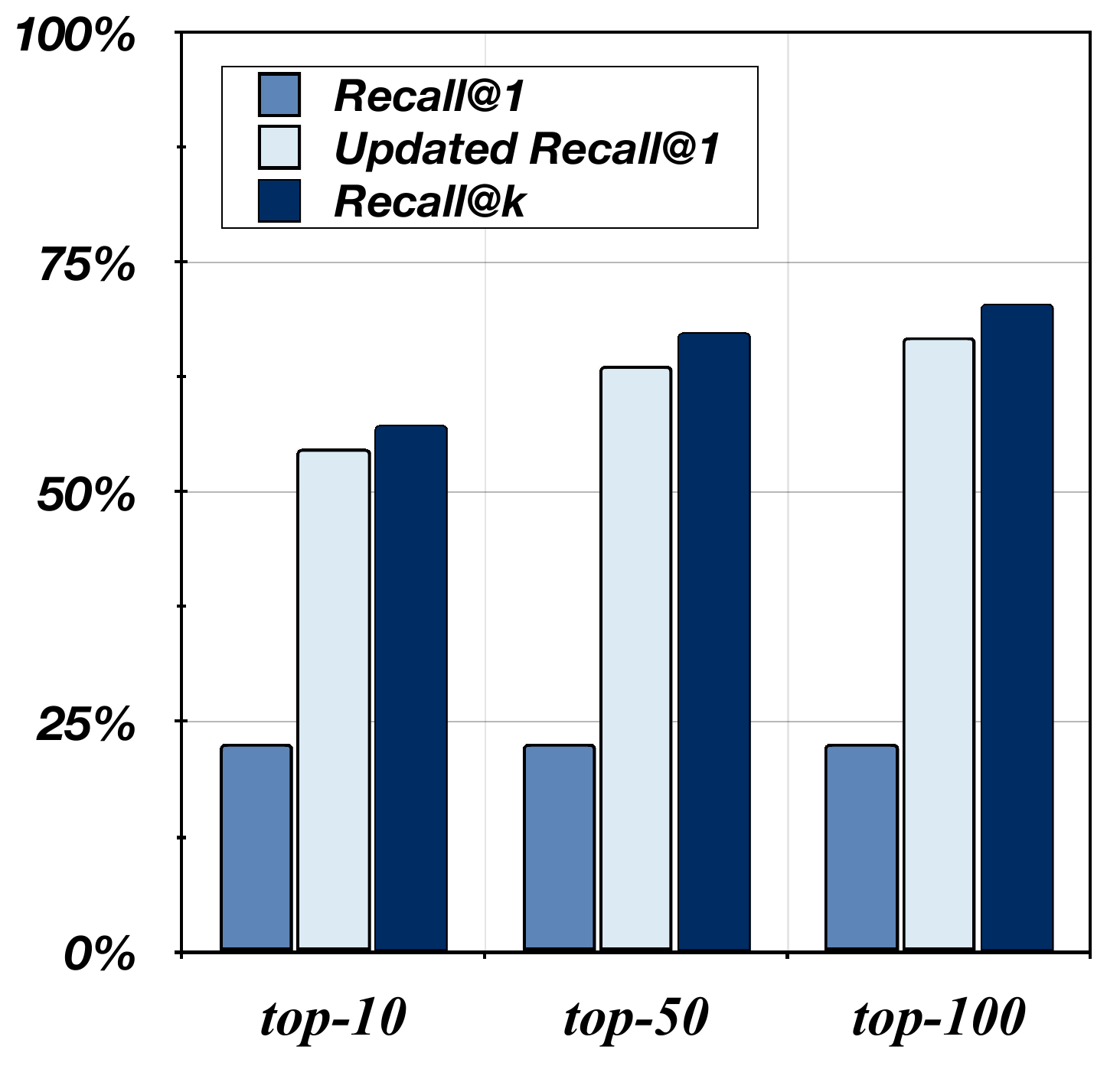}
        \caption{\binaryai{}}
        \label{fig:binaryai-match}
    \end{subfigure}
    \begin{subfigure}[t]{0.49\columnwidth}
        \centering
        \includegraphics[width=\textwidth]{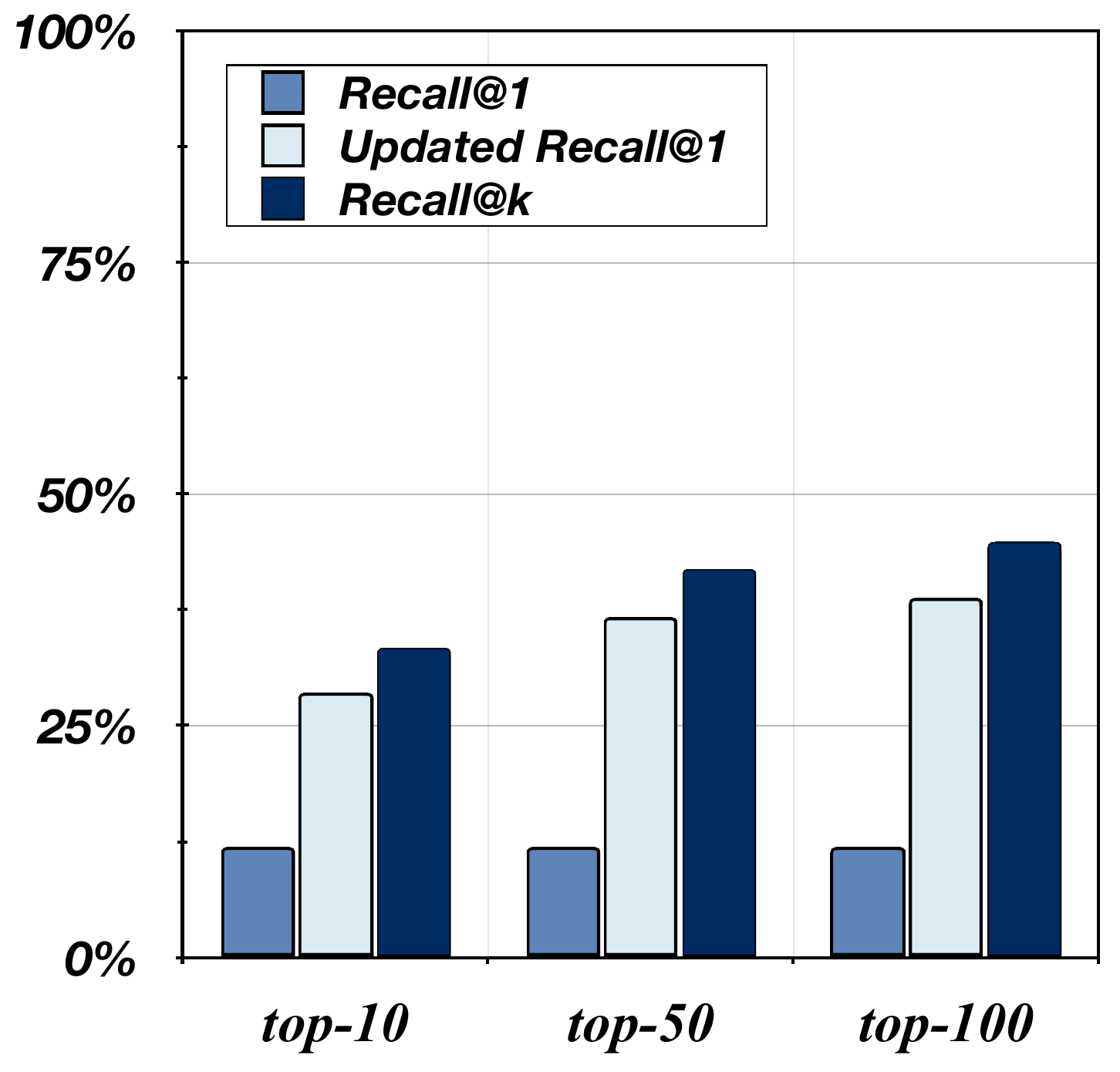}
        \caption{\codecmr{}}
        \label{fig:codecmr-match}
    \end{subfigure}
    \caption{Contribution of locality-driven matching}
    \label{fig:match-res}
\end{figure}

Then we further investigate the contribution of \approachlocality{} to binary source code matching. Specifically, we apply each newly matched source function from the phase of \approachlocality{} to update the corresponding \textit{top-1} similar function from \approachembedding{}. Figure~\ref{fig:match-res} presents the original \textit{recall@1} based on \approachembedding{}, the updated \textit{recall@1} by adding the newly matched source functions with different \textit{top-k} retrieved results as input of \approachlocality{}, and the corresponding \textit{recall@k} which means the upper bound of recall restricted by the capability of the model. Overall, the newly matched results by \approachlocality{} significantly improve the original \textit{recall@1} that almost reaches the upper bound for both \binaryai{} and \codecmr{}. For \binaryai{}, \approachlocality{} increases the \textit{recall@1} from 22.73\% to 54.70\% with the upper bound as 57.35\% for \textit{top-10}, and further increase the \textit{recall@1} to 66.90\% with the upper bound as 70.45\% for \textit{top-100}. Similarly, \approachlocality{} increases the \textit{recall@1} from 11.92\% to 28.61\% with the upper bound as 33.46\% for \textit{top-10} in \codecmr{}, indicating that as long as the model retrieves the exactly matched source function within the \textit{top-k} results, \approachlocality{} can effectively identify and update the matched source function as the new \textit{top-1} result. Such a result is promising for improving binary source code matching, indicating that we can focus on enhancing the model capability of \textit{top-k} retrieval in the future and leverage \approachlocality{} to further identify the matched results. Overall, we can demonstrate the effectiveness of the two-phase design of binary source code matching to capture both syntactic and semantic code features in \binaryai{}.

\mybox{\textbf{Finding 4:} Locality-driven matching significantly increases the overall accuracy of binary source code matching, facilitating the downstream binary SCA.}

\subsubsection{RQ3: Accuracy of TPL Detection}
\label{sec:eval-rq3}

\begin{figure}[htbp]
    \centering
    \includegraphics[width=0.92\columnwidth]{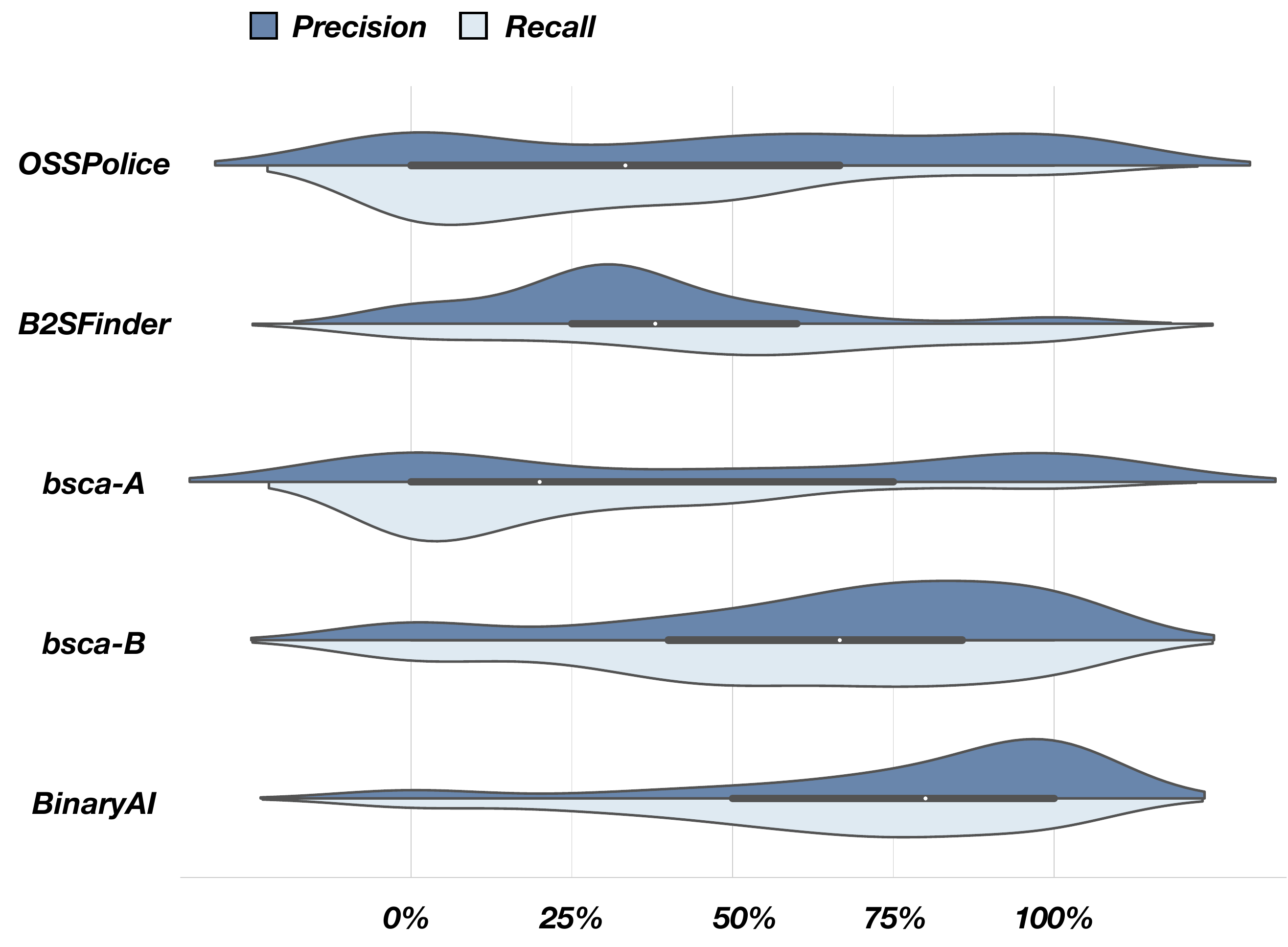}
    \caption{Distribution of binary SCA results}
    \label{fig:sca-res}
\end{figure}

Lastly, we compare the performance of \binaryai{} with the existing tools in terms of binary-to-source SCA. Table~\ref{tab:smart-sca} demonstrates the overall result of TPL detection for \scalabel{} labeled components within \scabin{} binary files. We can observe that, in general, \binaryai{} significantly outperforms all the other SCA tools. For instance, \binaryai{} significantly outperforms typical academic binary-to-source SCA techniques \osspolice{} 
and \bsfinder{}. 
Furthermore, \binaryai{} can even outperform well-recognized commercial binary SCA product \bdba{} (85.84\% vs. 73.36\% precision, 64.98\% vs. 59.81\% recall, and 73.97\% vs. 65.90\% \textit{F1}). Figure~\ref{fig:sca-res} presents the distribution of precision and recall for TPL detection across the \scabin{} binary files. We observe that \binaryai{} can dominate the precision and recall of component identification in most binary files, followed by the \bdba{}. On the contrary, \osspolice{}, \bsfinder{}, and \scantist{} cannot generalize well to our SCA test cases with compromised precision or recall.

\parabf{False Positive Analysis.} We investigate 112 false positives from all 791 identified components which is rather limited in the domain of binary-to-source SCA, and find that all of them are related to the limitation of the TPL dependency. Specifically, there are overlapped function features between false positives and the correct TPLs, while we fail to filter out the false positives based on the TPL dependency generated by \tplite{}. 

\parabf{False Negative Analysis.} We investigate all the false negatives, where most of them (312 out of 366) are caused by the partial reuse of the third-party components. In particular, the binary file only reuses a small fraction of functions from the labeled TPL, leading to a lower ratio than the pre-defined threshold $\theta$. For instance, \textit{nano\_node} only reuses 8 functions from \textit{leveldb} that causes the false positive. Note that the partial TPL reuse is generally the challenge of SCA~\cite{woo2021centris, tplite, ossfp}, and other reasons for false negatives include missing the corresponding source functions due to decompilation errors and the capability of the model to retrieve similar functions. 

\mybox{\textbf{Finding 5:} \binaryai{} dominates the performance of TPL detection among the state-of-the-art binary SCA tools.}

\begin{table}[t]
    \centering
    \caption{Result of binary-to-source SCA}
    \label{tab:smart-sca}
    \setlength\tabcolsep{12pt}
    \renewcommand\arraystretch{1.05}
    \begin{adjustbox}{width=0.98\columnwidth}
    \begin{tabular}{lcccccc}
    \toprule[1.2pt]
    \multirow{2}{*}{\textbf{Tool}} & \multicolumn{6}{c}{\textbf{Verification of TPL Detection}} \\
    \cmidrule{2-7}
        & \#TP  & \#FP  & \#FN   &  P (\%)  & R (\%) & F1 (\%) \\   
    \midrule
    \osspolice{} & \multicolumn{1}{|c}{348} & 191 & 697 & 64.56 & 33.30 & 43.94 \\
    \bsfinder{}  & \multicolumn{1}{|c}{574} & 1232 & 471 & 31.78 & 54.93 & 40.26 \\
    \scantist{}  & \multicolumn{1}{|c}{232} & 108 & 813 & 68.24 & 22.20 & 33.50 \\
    \bdba{}      & \multicolumn{1}{|c}{625} & 227 & 420 & 73.36 & 59.81 & 65.90 \\
    \midrule
    \textbf{\binaryai{}}  & \multicolumn{1}{|c}{\textbf{679}} & \textbf{112} & \textbf{366} & \textbf{85.84} & \textbf{64.98} & \textbf{73.97} \\
    \bottomrule[1.2pt]
    \end{tabular}
    \end{adjustbox}
\end{table}

\section{Threats to Validity}

\noindent \textbf{Threats to internal validity.} The threat to internal validity mainly lies in the design of \binaryai{}. To reduce this threat, we have spent over two years exploring technical solutions, including training the model to directly learn structured code features (e.g., AST, control, or data flow information) as part of the embedding. However, their performance is far inferior to the current design of \binaryai{}. Additionally, to reduce the threat of implementation, we invite three senior engineers in the relevant domain to review the code and ensure its correctness and consistency carefully.

\noindent \textbf{Threats to external validity.} The threat to external validity mainly lies in the subjects and dataset. To reduce this threat, we compare the model in \binaryai{} with the state-of-the-art model \codecmr{} in the domain of binary source code matching. For the downstream SCA task, we compare \binaryai{} with four typical binary SCA tools from both industry and academia (\osspolice{}, \bsfinder{}). For the dataset, we follow previous works~\cite{woo2021centris, tang2022towards, wang2022jtrans} to construct substantial binary source function pairs as the training dataset by building an automatic compilation pipeline for ArchLinux packages.
Meanwhile, we collect a large-scale TPL dataset and create the largest corpus of source functions respectively in binary-to-source SCA and binary source code matching. Considering the lack of publicly available ground-truth data, it also takes the authors excessive manual effort to calibrate the ground-truth components and label the correspondence of functions.

\noindent \textbf{Threats to construct validity.} The threat to construct validity mainly lies in the adopted metrics in our evaluation. To reduce this threat, we strictly follow prior works~\cite{woo2021centris, tplite, yu2020codecmr, wang2022jtrans, tang2022libdb} to evaluate \binaryai{} with multiple widely-used metrics, i.e., \textit{MRR}, \textit{Recall@k}, \textit{Precision}, \textit{Recall}, and \textit{F1} score.

\section{Related Work}
\subsection{Software Composition Analysis}

Many existing SCA techniques based on binary analysis~\cite{jiang2023evaluating,wu2022one,wu2022evaluating,li2017steelix,wu2023enhancing,huang2020pangolin,stephens2016driller} employ various feature extraction approaches and matching algorithms to improve the accuracy of TPL detection. B2SFinder~\cite{yuan2019b2sfinder} extracts control-flow structures to capture the target program's semantic information and allocates weight to different features. ModX~\cite{yang2022modx} takes a modularization approach that clusters functions into semantically-based modules. Xu et al.~\cite{xu2021interpretation} propose a multi-level birthmark model that extracts program features on three levels to deal with obfuscation. Tang et al.~\cite{tang2022libdb} propose LibDB, which utilizes syntactic and function embedding features. It also filters out duplicated features with the assistance of function call graphs. OSSPolice~\cite{duan2017identifying} adopts a hierarchical indexing scheme to locate true matches. Multiple SCA works are designed to operate in a source-to-source setting. SourcererCC~\cite{sajnani2016sourcerercc} utilizes an optimized partial index and filtering heuristics to detect open-source code clones. Lopes et al.~\cite{lopes2017dejavu} further adopt SourcererCC to construct a duplicate code map called D{\'e}j{\`a}Vu for the code repositories on GitHub. Centris~\cite{woo2021centris} takes function signature as the basic feature and derives TPL dependencies based on function birth time to alleviate internal code clones. TPLite~\cite{tplite} utilizes hierarchical path information to identify the origin TPL and centrality analysis to filter out false positives. 

In addition to the C/C++ ecosystem, many SCA techniques are specifically designed to identify components of Android applications. ATVHunter~\cite{zhan2021atvhunter} takes a two-phase approach which uses control-flow graphs as features in the first stage and the opcode of control-flow graphs in the second stage. LibScout~\cite{backes2016reliable} employs class hierarchy information that does not rely on concrete code to improve resilience against code obfuscation. Zhang et al.~\cite{zhang2019libid} propose LibID to cope with a wider range of obfuscation scenarios in Android. It leverages class signatures as features and a three-stage matching scheme to ensure that a library exists. In this paper, we propose \binaryai{}, which is the first to adopt a transformer-based model in the domain of binary-to-source SCA. Additionally, \binaryai{} leverages \linklocality{} to enhance the accuracy of binary source code matching and the downstream SCA task.

\subsection{Code Clone Detection}

Code clone detection is widely used to evaluate code similarities among software projects. It can be divided into two categories: binary-to-source code matching and binary-to-binary code matching. In the binary-to-source matching level, CodeCMR~\cite{yu2020codecmr} adopts DPCNN for source code feature extraction and GNN for binary code feature extraction. 
Many existing works are in the binary-to-binary matching level. Asm2Vec~\cite{ding2019asm2vec} is an assembly code representation learning model, which produces a vector representation for each assembly function in the repository and uses cosine similarity to retrieve the top-k ranked candidates as results. InnerEye~\cite{Zuo2018NeuralMT} proposes a cross-lingual deep learning approach at the basic-block level. jTrans~\cite{wang2022jtrans} presents a transformer-based approach that uses a jump-aware representation of the analyzed binary code and a newly-designed pre-training task to embed the control flow information into the language model. Based on graph representation learning, Xu et al.~\cite{xu2017neural} propose Gemini, a neural network-based approach, which applies Structure2vec to embed the attributed control flow graph (ACFG) and then measures the distance between the embeddings for two functions. Kim et al.~\cite{kim2022improving} present XBA, a deep learning-based technique, which first abstracts binary disassembly graphs (BDGs), and then formulates the binary code representation learning as a graph alignment problem. DeepBinDiff~\cite{Li2019LearningPC} relies on both the code semantic information distilled by NLP techniques and program-wide control flow information to generate embeddings at the basic block level.
VulHawk~\cite{Luo2023VulHawkCV} proposes an intermediate representation function model with NLP techniques and graph convolutional networks to generate function embeddings and an entropy-based adaptor to alleviate the differences caused by various file environments in function embeddings.

\section{Conclusion}
In this paper, we propose a novel binary-to-source SCA technique, \binaryai{}, to alleviate the problem that existing binary-to-source SCA techniques suffer from redundancy in the large-scale TPL dataset and few syntactic features between reused TPLs and target binary files. \binaryai{} trains a transformer-based model to generate function embeddings by learning the token-based syntactic feature of the code language and leverages locality-driven matching to enrich semantic features for further identifying the positive samples. Based on the matched source functions, \binaryai{} performs SCA by detecting the reused TPL components in the target binary file. The evaluation results indicate that the embedding model significantly outperforms \codecmr{} with  22.54\% recall@1 and 0.34 \textit{MRR}. Additionally, \binaryai{} with \approachlocality{} can further improve the binary source code matching and its TPL detection result exceeds all state-of-the-art binary-to-source SCA tools.
\section*{Acknowledgement}
This work is partially supported by the National Natural Science Foundation of China (Grant No. 62372220). Ling Jiang would like to dedicate this paper to the love of his fiancée. 

\clearpage
\balance
\bibliographystyle{ACM-Reference-Format}
\bibliography{ref}

\end{document}